\documentclass[letterpaper, 11pt]{article}  

\usepackage{graphicx}
\usepackage{graphics}
\usepackage{amsmath}
\usepackage{amssymb}
\usepackage{xy}
\usepackage{array}                                                       

\title{\LARGE \bf
Robust $H_{\infty}$ Filter Design for Lipschitz Nonlinear Systems
via Multiobjective Optimization}
\date{ }

\author{Masoud Abbaszadeh$^{\dag}$$^{\ddagger}$\thanks{Author to whom correspondence should be addressed. Phone: +1-519-729-0741, Fax:+1-519-747-5284}\\
    masoud@ece.ualberta.ca
    \and
    Horacio J. Marquez$^\dag$\\
     marquez@ece.ualberta.ca
    }
\begin{document}
\maketitle \begin{center}\thanks{$^{\dag}$Department of Electrical
and Computer Engineering, University of Alberta, Edmonton, Alberta,
Canada, T6G 2V4 \\ $^{\ddagger}$Department of Research and Development, Maplesoft, Waterloo, Ontario, Canada, N2V 1K8}\end{center}


\begin{abstract}
In this paper, a new method of $H_{\infty}$ observer design for
Lipschitz nonlinear systems is proposed in the form of an LMI
optimization problem. The proposed observer has guaranteed decay
rate (exponential convergence) and is robust against unknown
exogenous disturbance. In addition, thanks to the linearity of the
proposed LMIs in the admissible Lipschitz constant, it can be
maximized via LMI optimization. This adds an extra important feature
to the observer, robustness against nonlinear uncertainty. Explicit
bound on the tolerable nonlinear uncertainty is derived. The new LMI
formulation also allows optimizations over the disturbance
attenuation level ($H_{\infty}$ cost). Then, the admissible
Lipschitz constant and the disturbance attenuation level of the
$H_{\infty}$ filter are simultaneously optimized through LMI
multiobjective optimization.
\end{abstract}


\section{Introduction}

    The design of nonlinear state observers has been an area of constant
research for the last three decades and as a result, a wide variety
of design techniques for nonlinear observers exist in the
literature. Despite important progress, many outstanding problems
still remain unsolved. A class of nonlinear systems of special
attention is the so-called Lipschitz systems in which the
mathematical model of the system satisfies a Lipschitz continuity
condition. Many practical systems satisfy the Lipschitz condition,
at least locally. Roughly speaking, in these systems, the rate of
growth of the trajectories is bounded by the rate of growth of the
states.  Observer design for Lipschitz systems was first considered
by Thau in his seminal paper \cite{Thau} where he obtained a
sufficient condition to ensure asymptotic stability of the observer.
Thau's condition provides a very useful analysis tool but does not
address the fundamental design problem. Encouraged by Thau's result,
several authors studied observer design for Lipschitz systems
\cite{Raghavan, Rajamani, Rajamani2, Aboky, Pertew}. All these
methods share a common structure for the error dynamics of the
nonlinear systems; namely the error dynamics can be represented as a
linear system with a sector bounded nonlinearity in feedback. This
type of problems are both theoretically and numerically tractable
because they can be formulated as convex optimization problems
\cite{Howell}, \cite{Boyd}. Raghavan formulated a procedure to
tackle the design problem. His algorithm is based on solving an
algebraic Riccati equation to obtain the static observer gain
\cite{Raghavan}. Unfortunately, Raghavan's algorithm often fails to
succeed even when the usual observability assumptions are satisfied.
Raghavan showed that the observer design might still be tractable
using state transformations. Another shortcoming of his algorithm is
that it does not provide insight into what conditions must be
satisfied by the observer gain to ensure stability. A rather
complete solution of these problems was later presented by Rajamani
\cite{Rajamani}. Rajamani obtained necessary and sufficient
conditions on the observer matrix that ensure asymptotic stability
of the observer error and formulated a design procedure, based on
the use of a gradient based optimization method. He also discussed
the equivalence between the stability condition and the minimization
of the $H_{\infty}$ norm of a system in the standard form. However,
he pointed out that the design problem is not solvable as a standard
$H_{\infty}$ optimization problem since the regularity assumptions
required in the $H_{\infty}$ framework are not satisfied. Using
Riccati based approach, Pertew et. al. \cite{Pertew} showed that the
condition introduced in \cite{Rajamani} is related to a modified
$H_{\infty}$ norm minimization problem satisfying all of the
regularity assumptions. It is worth mentioning that the $H_{\infty}$
problem in \cite{Rajamani} is associated with the nominal stability
of the observer error dynamics while no disturbance attenuation is
considered. Moreover, in all of the above references, the system
model is assumed to be perfectly known with no uncertainty or
disturbance.  In order to guarantee robustness against unknown
exogenous disturbance, the nonlinear $H_{\infty}$ filtering was
introduced by De Souza et. al. \cite{deSouza1, deSouza2} via the
Riccati approach. In an $H_{\infty}$ observer, the
$\mathcal{L}_{2}$-induced gain from the norm-bounded exogenous
disturbance signals to the observer error is guaranteed to be below
a prescribed level. On the other hand, the restrictive regularity
assumptions in the Riccati approach can be relaxed using linear
matrix inequalities (LMIs). In this paper, we introduce a novel
nonlinear $H_{\infty}$ observer design method for Lipschitz
nonlinear systems based on the LMI framework. Our solution follows
the same approach as the original problem of Thau and proposes a
natural way to tackle the problem, directly. Unlike the methods of
\cite{Raghavan, Rajamani, Pertew}, the proposed LMIs can be
efficiently solved using commercially available software without any
tuning parameters. In all aforementioned references, the Lipschitz
constant of the system is assumed to be known and fixed. In this
paper, the resulting LMIs are formulated such that to be linear in
the Lipschitz constant of the nonlinear system. This adds an
important extra feature to the observer, robustness against
nonlinear uncertainty. Maximizing the admissible Lipschitz constant,
the observer can tolerate some nonlinear uncertainty for which an
explicit norm-wise bound is derived. In addition to this robustness,
we will extend our result such that the observer disturbance
attenuation level (the $H_{\infty}$ feature of the observer) can be
optimized as well. Then, both the admissible Lipschitz constant and
the disturbance attenuation level are optimized simultaneously
through multiobjective convex optimization. The rest of the paper is
organized as follows: Section 2, introduces the problem and some
background. In Section 3, the LMI formulation of the problem and our
observer design algorithm are proposed. The observer guaranteed
decay rate and robustness against nonlinear uncertainty are
discussed. In Section 4, we expand the result of Section 3, to an
$H_{\infty}$ nonlinear observer design method. Section 5, is devoted
to the simulators optimization of the observer features through
multiobjective optimization. In section 6, the proposed observer
performance is shown in some illustrative examples.

\section{Preliminaries and Problem Statement}

Consider the following continuous-time nonlinear system
\begin{align}
\dot{x}(t)&=Ax(t)+ \Phi(x,u)\hspace{7mm} A \in\mathbb{R}^{n\times
n}\label{con1}\\
y(t)&=Cx(t)\hspace{23mm} C \in\mathbb{R}^{n\times p}\label{con2}
\end{align}
where $x\in {\mathbb R} ^{n} ,u\in {\mathbb R} ^{m} ,y\in {\mathbb
R} ^{p} $ and $\Phi(x,u)$ contains nonlinearities of second order or
higher. We assume that the system (\ref{con1})-\eqref{con2} is
locally Lipschitz in a region $\mathcal{D}$ including the origin
with respect to $x$, uniformly in $u$, i.e.:
\begin{eqnarray}
\|\Phi(x_{1},u^{*})-\Phi(x_{2},u^{*})\|\leqslant\gamma\|x_{1}-x_{2}\|
\hspace{7mm}\forall \, x_{1} (k),x_{2} (k)\in \mathcal{D}
\end{eqnarray}
where $\|.\|$ is the induced 2-norm, $u^{*}$ is any admissible
control signal and $\gamma>0$ is called the Lipschitz constant. If
the nonlinear function $\Phi$ satisfies the Lipschitz continuity
condition globally in $\mathbb{R}^{n}$, then the results will be
valid globally. Consider now an observer of the following form
\begin{align}
\dot{\hat{x}}(t)=A\hat{x}(t)+\Phi(\hat{x},u)+L(y-C\hat{x})\label{observer1}.
\end{align}

The observer error dynamics is given by
\begin{align}
e(t)&\triangleq x(t)-\hat{x}(t)
\\\dot{e}(t)&=(A-LC)e(t)+\Phi(x,u)-\Phi(\hat{x},u).\label{error1}
\end{align}

The goal is to find a gain, $L$, such that:
\begin{itemize}
  \item In the absence of disturbance, the observer error dynamics is asymptotically stable i.e.: $\lim_{t\rightarrow \infty}
e(t)=0$.
  \item In the presentence of unknown exogenous disturbance, a disturbance attenuation level is
guaranteed. ($H_{\infty}$ performance).
\end{itemize}

The result is simple and yet efficient with no regularity
assumption. The observer error dynamics is asymptotically stable
with guaranteed decay rate (the convergence is actually exponential
as we will see). In addition, the observer is robust against
nonlinear uncertainty and exogenous disturbance. The dismissible
Lipschitz constant which as will be shown, determines the robustness
margin against nonlinear uncertainty, and the disturbance
attenuation level (the $H_{\infty}$ cost), are optimized through LMI
optimization.

\addtolength{\textheight}{-3cm}   

\section{An Algorithm for Nonlinear Observer Design}

In this section an LMI approach for the nonlinear observer design
problem introduced in Section 2 is proposed and some performance
measures of the observer are optimized.

\subsection{Maximizing the Admissible Lipschitz Constant}
We want to maximizes the admissible Lipschitz constant of the
nonlinear system (1)-(2) for which the observer error dynamics is
asymptotically stable. The following theorem states the main result of this section.\\

\emph{\textbf{Theorem 1.} Consider the Lipschitz nonlinear system
(\ref{con1})-(\ref{con2}) along with the observer (\ref{observer1}).
The observer error dynamics (\ref{error1}) is (globally)
asymptotically stable with maximum admissible Lipschitz constant if
there exist scalers $\epsilon
> 0$ and $\xi > 0$ and matrices $P > 0$ and F such that the
following LMI optimization problem has a solution. }
\begin{align}
\hspace{-2cm}min (\xi) \notag
\end{align}
\hspace{4cm}\emph{s.t.}
\begin{align}
&A^{T}P+PA-C^{T}F^{T}-FC < -I-\epsilon I  \label{LMI1}
\\&\left[
  \begin{array}{cc}
    \frac{1}{2}\xi I & P \\
    P &  \frac{1}{2}\xi I
  \end{array}
\right]>0 \label{LMI2}
\end{align}
\emph{once the problem is solved}
\begin{align}
L&=P^{-1}F \label{L1}
\\\gamma^{*}&\triangleq\max(\gamma)=\xi^{-1}
\end{align}

\textbf{Proof:} Suppose $Q=I$. The original problem as discussed in
section 2, can be written as
%

\begin{equation}
\hspace{-5cm} \min (\lambda_{max}(P)) \notag
\end{equation}
\hspace{3cm} s.t.
\begin{align}
(A-LC)^{T}P+P^{T}(A-LC)&=-I\label{lyap2}\\
1-2\gamma.\lambda_{max}(P)&>0\label{cond2}
\\ P&>0
\end{align}

which is a nonlinear optimization problem, hard to solve if not
impossible. We proceed by converting it into an LMI form. A
sufficient condition for existence of a solution for (\ref{lyap2})
is
\begin{equation}
\\\exists\hspace{1mm}\epsilon>0,
\hspace{1mm}(A-LC)^{T}P+P^{T}(A-LC)<-I-\epsilon I.
\end{equation}

The above can be written as
\begin{equation}
\\A^{T}P+PA-C^{T}L^{T}P-PLC<-I-\epsilon I
\end{equation}
which is a bilinear matrix inequality. Defining the new variable
\begin{equation}
\\F\triangleq PL\rightarrow L^{T}P^{T}=L^{T}P=F^{T}
\end{equation}
it becomes
\begin{equation}
\\A^{T}P+PA-C^{T}F^{T}-FC < -I-\epsilon I
\end{equation}
In addition, since $P$ is positive definite
$\bar{\sigma}(P)=\lambda_{max}(P)$. So, from (\ref{cond2}) we have
\begin{equation}
\\\bar{\sigma}(P)<\frac{1}{2\gamma}\label{cond3}
\end{equation}
which is equivalent to
\begin{equation}
\\(\frac{1}{2\gamma})^{2}I-P^{T}P>0
\end{equation}

using Schur's complement lemma

\begin{equation}
\left[
  \begin{array}{cc}
    \frac{1}{2\gamma}I & P \\
    P & \frac{1}{2\gamma}I \\
  \end{array}
\right]>0 \label{cond4}
\end{equation}

defining $\xi=\frac{1}{\gamma}$, (\ref{LMI2}) is achieved. $\triangle$\\

\emph{\textbf{Proposition1.}} {\emph{Suppose the actual Lipschitz
constant of the system is $\gamma$ and the maximum admissible
Lipschitz constant achieved by Theorem 2, is $\gamma^{*}$. Then, the
observer designed based on Theorem 2, can tolerate any additive
Lipschitz nonlinear uncertainty with Lipschitz constant less than or
equal $\gamma^{*}-\gamma$}}.\\

\textbf{Proof:} Assume a nonlinear uncertainty as follows
\begin{align}
\Phi_{\Delta}(x,u)&=\Phi(x,u)+\Delta\Phi(x,u)
\\\dot{x}(t)&= Ax(t) + \Phi_{\Delta}(x,u)
\end{align}
where
\begin{align}
\|\Delta\Phi(x_{1},u)-\Delta\Phi(x_{2},u)\|\leqslant\Delta\gamma\|x_{1}-x_{2}\|.
\end{align}
Based on Schwartz inequality, we have
\begin{eqnarray}
\|\Phi_{\Delta}(x_{1},u)-\Phi_{\Delta}(x_{2},u)\|&\leq&\notag
\|\Phi(x_{1},u)-\Phi(x_{2},u)\|+\|\Delta\Phi(x_{1},u)-\Delta\Phi(x_{2},u)\|\notag
\\&\leq& \gamma\|x_{1}-x_{2}\|+\Delta\gamma\|x_{1}-x_{2}\|.
\end{eqnarray}
According to the Theorem 1, $\Phi_{\Delta}(x,u)$ can be any
Lipschitz nonlinear function with Lipschitz constant less than or
equal to $\gamma^{*}$,
\begin{equation}
\|\Phi_{\Delta}(x_{1},u)-\Phi_{\Delta}(x_{2},u)\|\leq\gamma^{*}\|x_{1}-x_{2}\|
\end{equation}
so, there must be
\begin{eqnarray}
\gamma+\Delta\gamma\leq\gamma^{*}\rightarrow\Delta\gamma\leq\gamma^{*}-\gamma.
\ \ \ \triangle
\end{eqnarray}

\emph{\textbf{Remark 1.}} If one wants to design an observer for a
given system with known Lipschitz constant, then the LMI
optimization problem can be reduced to an LMI feasibility problem
(just satisfying the constraints) which is easier.\\

>From Theorem 2, it is clear that the gain $L$ obtained via solving
the LMI optimization problem, can lead to stable error dynamics for
every member in the class of the Lipschitz nonlinear functions with
Lipschitz constant less than or equal to $\gamma^{*}$. Thus, it
neglects the structure of the given nonlinear function. It is
possible to take advantage of the structure of the $\Phi(x,u)$ in
addition to the fact that its Lipschitz constant is $\gamma$.
According to Proposition 1, the margin of robustness against
nonlinear uncertainty is $\gamma^{*}-\gamma$. The Lipschitz constant
of the systems can be reduced using appropriate coordinates
transformations. The transformation matrices that are picked are
problem specific and they reflect the structure of the given
nonlinearity \cite{Raghavan}. The robustness margin can then be
modified through coordinates transformations.
Finding the Lipschitz constant of a function is itself a global
optimization problem, since the Lipschitz constant is the supremum
of the magnitudes of directional derivatives of the function as
shown in \cite{Khalil} and \cite{Marquez}. If the analytical form of
the nonlinear function and its derivatives are known explicitly, any
appropriate global optimization method may be applied to find the
Lipschitz constant. If only the function values can be evaluated, a
stochastic random search and probability density function fitting
method may be used \cite{Wood}.

\subsection{Guaranteed Decay Rate}

The decay rate of the system (\ref{error1}) is defined to be the
largest $\beta>0$ such that
\begin{eqnarray}
\lim_{t\rightarrow\infty} \exp(\beta t)\|e(t)\|=0
\end{eqnarray}
holds for all trajectories $e$. We can use the quadratic Lyapunov
function $V (e)=e^{T}Pe$ to establish a lower bound on the decay
rate of the (\ref{error1}). If $\frac{dV(e(t))}{dt}\leqslant-2\beta
V (e(t))$ for all trajectories, then $V(e(t)) \leqslant \exp(-2\beta
t)V(e(0))$, so that $\|e(t)\|\leqslant \exp(-\beta
t)\kappa(P)^{\frac{1}{2}}\|e(0)\|$ for all trajectories, where
$\kappa(P)$ is the condition number of P and therefore the decay
rate of the (\ref{error1}) is at least $\beta$, \cite{Boyd}. In
fact, decay rate is a measure of
observer speed of convergence.\\

\emph{\textbf{Theorem 3.} Consider Lipschitz nonlinear system
(\ref{con1})-(\ref{con2}) along with the observer (\ref{observer1}).
The observer error dynamics (\ref{error1}) is (globally)
asymptotically stable with maximum admissible Lipschitz constant and
guaranteed decay rate $\beta$, if there exist a fixed scaler $\beta>
0$, scalers $\epsilon> 0$ and $\xi > 0$ and matrices $P > 0$ and F
such that the following LMI optimization problem has a solution.}
\begin{align}
&\hspace{2cm} min (\xi) \notag
\\ s.t. \notag
\\&A^{T}P+PA+2\beta P-C^{T}F^{T}-FC < -I-\epsilon I\label{lyap4}
\\&\left[
  \begin{array}{cc}
    \frac{1}{2}\xi I & P \\
    P &  \frac{1}{2}\xi I
  \end{array}
\right]>0
\end{align}
\emph{once the problem is solved}
\begin{align}
L&=P^{-1}F
\\\gamma^{*}&\triangleq\max(\gamma)=\xi^{-1}
\end{align}

\textbf{Proof:} Consider the following Lyapunov function candidate
\begin{eqnarray}
V(t)=e^{T}(t)Pe(t)
\end{eqnarray}
then
\begin{equation}
\dot{V}(t)=\dot{e}^{T}(t)Pe(t)+e^{T}(t)P\dot{e}(t)=-e^{T}Qe+2e^{T}P(\Phi(x,u)-\Phi(\hat{x},u))^{T}\label{V1}.
\end{equation}
To have $\dot{V}(t)\leqslant-2\beta V(t)$ it suffices (\ref{V1}) to
be less than zero, where:
\begin{equation}
(A-LC)^{T}P+P^{T}(A-LC)+2\beta P=-Q \label{lyap3}.
\end{equation}
The rest of the proof is the same as the proof of Theorem 2.
$\Delta$

\section{Robust $H_{\infty}$ Nonlinear Observer}

In this section we extend the result of the previous section into a
new nonlinear robust $H_{\infty}$ observer design method. Consider
the system
\begin{eqnarray}
\dot{x}(t)&=& Ax(t) + \Phi(x,u)+B w(t)\hspace{4mm}\label{sys3}
\\ y(t)&=& Cx(t)+D w(t)\label{sys4}
\end{eqnarray}
where $w(t)\in\mathfrak{L}_{2}[0,\infty)$ is an unknown exogenous disturbance. suppose that
\begin{equation}
z(t)=He(t)
\end{equation}
stands for the controlled output for error state where $H$ is a
known matrix. Our purpose is to design the observer parameter $L$
such that the observer error dynamics is asymptotically stable and
the following specified $H_{\infty}$ norm upper bound is
simultaneously guaranteed.
\begin{equation}
\|z\|\leq\mu\|w\|.
\end{equation}
The following theorem introduces a new method for nonlinear robust
$H_{\infty}$ observer design. we first present an inequality
that will be used in the proof of our result.\\

\emph{\textbf{Lemma 1 \cite{Xu2}}. For any $x,y\in\mathbb{R}^{n}$
and any positive definite matrix $P\in\mathbb{R}^{n\times{n}}$, we
have}
\begin{equation}
2x^{T}y\leq x^{T}Px+y^{T}P^{-1}y
\end{equation}

\emph{\textbf{Theorem 4.} Consider the Lipschitz nonlinear system
(\ref{sys3})-(\ref{sys4}) with given Lipschitz constant $\gamma$,
along with the observer (\ref{observer1}). The observer error
dynamics is (globally) asymptotically stable with decay rate $\beta$
and minimum $\mathfrak{L}_{2}(w \rightarrow e)$ gain, $\mu$, if
there exist fixed scaler $\beta>0$, scalers $\alpha>1$, $\epsilon>
0$ and $\zeta>0$ and matrices $P>0$ and $F$ such that the
following LMI optimization problem has a solution.}\\
\begin{equation}
\hspace{-6cm} \ min (\zeta) \notag
\end{equation}
\hspace{3cm}\emph{s.t.}
\begin{align}
&A^{T}P+PA+2\beta P-C^{T}F^{T}-FC < -\alpha I-\epsilon I\label{LMI8}\\
&\left[
  \begin{array}{cc}
    \frac{1-\bar{\sigma}^{2}(H)}{2\gamma} I & P \\
    P &  \frac{1-\bar{\sigma}^{2}(H)}{2\gamma} I
  \end{array}
\right]>0 \label{LMI4} \\
&\left[
    \begin{array}{cc}
        H^{T}H+\frac{1}{2}(\gamma+\frac{1}{\gamma}-2\alpha)I &  PB-FD \\
        \\B^{T}P-D^{T}F^{T}  & -\zeta I \\
    \end{array}%
\right]<0 \label{LMI3}
\end{align}
\emph{Once the problem is solved}
\begin{align}
L&=P^{-1}F
\\\mu^{*}&\triangleq\min(\mu)=\sqrt{\zeta}
\end{align}

\textbf{Proof:} The observer error dynamics will be
\begin{eqnarray}
\dot{e}(t)=(A-LC)e(t)+\Phi(x,u)-\Phi(\hat{x},u)+(B-LD)w
\end{eqnarray}
consider the following Lyapunov function candidate
\begin{eqnarray}
V(t)=e^{T}(t)Pe(t)
\end{eqnarray}
then
\begin{equation}
\dot{V}(t)=\dot{e}^{T}(t)Pe(t)+e^{T}(t)P\dot{e}(t)=-e^{T}Qe\notag
\end{equation}
\begin{equation}
+2e^{T}P(\Phi(x,u)-\Phi(\hat{x},u))^{T}+e^{T}(PB-FD)w+w^{T}(B^{T}P-D^{T}F^{T})e
\end{equation}\\
where, $Q$ is as in (\ref{lyap3}). We select $Q=\alpha I$. If $w=0$
the error dynamics is as Theorem 2, so the LMIs (\ref{LMI1}) and
(\ref{LMI2}) which for $Q=\alpha I$ will become
\begin{align}
A^{T}P+PA+2\beta P-C^{T}F^{T}-FC &< -\alpha I-\epsilon I
\label{LMI5}
\\\left[
  \begin{array}{cc}
    \frac{\alpha}{2\gamma} I & P \\
    P &  \frac{\alpha}{2\gamma} I
  \end{array}
\right]&>0 \label{LMI6}
\end{align}
are sufficient for the asymptotic stability of the error dynamics. Having $\alpha>1$, (\ref{cond3}) always implies (\ref{LMI6}).\\
\indent Based on Rayleigh inequality
\begin{equation}
e^{T}Qe\leq\lambda_{max}(Q)e^{T}e \label{ineq1}
\end{equation}
\indent Using Lemma 1 we can write
\begin{equation}
2e^{T}P(\Phi(x,u)-\Phi(\hat{x},u)) \leq
e^{T}Pe+(\Phi(x,u)-\Phi(\hat{x},u))^{T}PP^{-1}P(\Phi(x,u)-\Phi(\hat{x},u))\notag
\end{equation}
\begin{equation}
=e^{T}Pe+(\Phi(x,u)-\Phi(\hat{x},u))^{T}P(\Phi(x,u)-\Phi(\hat{x},u))\label{ineq2}
\end{equation}
based on Rayleigh inequality we have
\begin{equation}
\|e^{T}Pe\|\leq \lambda_{max}(P)\|e\|^{2}=\lambda_{max}(P)e^{T}e
\end{equation}
\begin{equation}
\|(\Phi(x,u)-\Phi(\hat{x},u))^{T}P(\Phi(x,u)-\Phi(\hat{x},u))\|\leq\lambda_{max}(P)\|\Phi(x,u)-\Phi(\hat{x},u)\|^{2}\notag
\end{equation}
\begin{equation}
\leq\gamma^{2}\lambda_{max}(P)\|e\|^{2}=\gamma^{2}\lambda_{max}(P)e^{T}e
\end{equation}
therefore, from the above and (\ref{cond3}),
\begin{equation}
2e^{T}P(\Phi(x,u)-\Phi(\hat{x},u))\leq
(1+\gamma^{2})\lambda_{max}(P) e^{T}e\leq
\frac{1}{2}(\gamma+\frac{1}{\gamma})e^{T}e \label{ineq3}.
\end{equation}
According to (\ref{ineq1}) and (\ref{ineq3}) and knowing that
$Q=\alpha I$, we have
\begin{eqnarray}
\dot{V}(t)\leq\frac{1}{2}(\gamma+\frac{1}{\gamma}-2\alpha)e^{T}e+e^{T}(PB-FD)w+w^{T}(B^{T}P-D^{T}F^{T})e.
\end{eqnarray}
\indent Now, we define
\begin{equation}
J=\int^{\infty}_{0}(z^{T}z-\zeta w^{T}w) dt
\end{equation}
therefore
\begin{equation}
J<\int^{\infty}_{0}(z^{T}z-\zeta w^{T}w+\dot{V}) dt
\end{equation}
it follows that a sufficient condition for $J\leq0$ is that
\begin{equation}
\forall t\in[0,\infty),\hspace{5mm} z^{T}z-\zeta
w^{T}w+\dot{V}\leq0
\end{equation}
but we have
\begin{equation}
z^{T}z-\zeta w^{T}w+\dot{V}\leq e^{T}H^{T}He-\zeta w^{T}w+\dot{V}
e^{T}H^{T}He+\frac{1}{2}(\gamma+\frac{1}{\gamma}-2\alpha)e^{T}e\notag\notag
\end{equation}
\begin{equation}
+e^{T}(PB-FD)w+w^{T}(B^{T}P-D^{T}F^{T})e-\zeta w^{T}w=\notag
\end{equation}
\begin{eqnarray}
\left[
\begin{array}{c}
  e \\
  w
\end{array}
\right]^{T}
 \left[
    \begin{array}{cc}
        {H^{T}H}+\frac{1}{2}(\gamma+\frac{1}{\gamma}-2\alpha)I  &  PB-FD \\
        \\B^{T}P-D^{T}F^{T}  & -\zeta I \\
    \end{array}%
\right] \left[
\begin{array}{c}
  e \\
  w \\
\end{array}
\right]\label{ineq4}.
\end{eqnarray}
Thus, a sufficient condition for $J\leq0$ is that the above matrix
which is the same as (\ref{LMI3}) be negative definite. Then
\begin{equation}
z^{T}z-\zeta w^{T}w\leq0\rightarrow\|z\|\leq\sqrt{\zeta}\|w\|
\end{equation}
\indent Up until now, we have the LMIs (\ref{LMI5}), (\ref{cond4})
and (\ref{LMI3}). If these LMIs are all feasible, then the problem
is solvable and the observer synthesis is complete. However,
(\ref{cond4}) can be slightly modified to improve its feasibility. We proceed as follows:\\
\indent Inequality (\ref{ineq2}) can be rewritten as follows
\begin{equation}
2e^{T}P(\Phi(x,u)-\Phi(\hat{x},u))\leq\
2\gamma\lambda_{max}(P)e^{T}e
\end{equation}
following the same steps, the matrix in (\ref{ineq4}) will become
\begin{eqnarray}
 \left[
    \begin{array}{cc}
        H^{T}H+[2\gamma\lambda_{max}(P)-\alpha]I &  PB-FD \\
        \\B^{T}P-D^{T}F^{T}  & -\zeta I \\
    \end{array}%
\right] <0. \label{ineq5}
\end{eqnarray}
The above matrix can not be used together with (\ref{LMI5}) and
(\ref{LMI6}) because it includes $P$ as one of the LMI variables,
thus resulting in a problem that is not linear in $P$. It can,
however, give us another insight about $\lambda_{max}(P)$. According
to the Schur's complement lemma, (\ref{ineq5}) is equivalent to
\begin{equation}
-\zeta I < 0
\end{equation}
\begin{equation}
H^{T}H+[2\gamma\lambda_{max}(P)-\alpha]I+\frac{1}{\zeta}(PB-FD)(PB-FD)^{T}<0.
\end{equation}
The third term in the above is always nonnegative, so it is
necessary to have
\begin{equation}
H^{T}H+[2\gamma\lambda_{max}(P)-\alpha]I<0 \label{ineq6}
\end{equation}
but as for any other symmetric matrix, for $H^{T}H$, we have
\begin{equation}
\lambda_{min}(H^{T}H)I\leq H^{T}H\leq\lambda_{max}(H^{T}H)I
\end{equation}
or according to the definition of singular values
\begin{equation}
\underline{\sigma}^{2}(H)I\leq H^{T}H\leq\ \bar{\sigma}^{2}(H)I
\end{equation}
therefore, a sufficient condition for (\ref{ineq6}) is
\begin{equation}
\bar{\sigma}^{2}(H)+2\gamma\lambda_{max}(P)-\alpha<0
\end{equation}
or
\begin{equation}
\lambda_{max}(P)<\frac{\alpha-\bar{\sigma}^{2}(H)}{2\gamma}\label{LMI7}
\end{equation}
but (\ref{cond3}) must be also  satisfied. To have both
(\ref{cond3}) and (\ref{LMI7}), it is sufficient that
\begin{equation}
\lambda_{max}(P)<\frac{1-\bar{\sigma}^{2}(H)}{2\gamma}
\end{equation}
which is equivalent to (\ref{LMI4}). $\triangle$ \\

\emph{\textbf{Remark 2.}} Similar to Remark 1, if one wants to
design an observer for a given system with known Lipschitz constant
and with a prespecified $\mu$, the LMI optimization problem is
reduced to an LMI feasibility problem.\\

%
%

\emph{\textbf{Remark 3.}} As an additional opportunity, we can first
maximize the admissible Lipschitz constant using Theorem 3, and then
minimize $\mu$ for the maximized $\gamma$, using Theorem 4. In this
case, according to Proposition 1, robustness against nonlinear
uncertainty is also guaranteed. In the next section, we will show
that how $\gamma$ and $\mu$ can be simultaneously optimized using
convex multiobjective optimization. It is clear that if no decay
rate is specified, then the term $2\beta P$ will be eliminated from
LMI (\ref{LMI8}) in Theorem 4.


\section{Combined Performance using Multiobjective Optimization}

The LMIs proposed in Theorem 4 are linear in both admissible
Lipschitz constant and disturbance attenuation level and as
mentioned earlier, each can be optimized. A more realistic problem
is to choose the observer gain matrix by combining these two
performance measures. This leads to a Pareto multiobjective
optimization in which the optimal point is a trade-off between two
or more linearly combined optimality criterions. Having a fixed
decay rate, the optimization is over $\gamma$ (maximization) and
$\mu$ (minimization), simultaneously. The following theorem is in
fact a generalization of the results of \cite{Raghavan, Rajamani,
Rajamani2, Aboky, Pertew, Zhu}, and  \cite{deSouza1} (for systems of
class \eqref{con1}-\eqref{con2}) in which the Lipschitz constant is
assumed to be known and fixed and the result of \cite{Howell} in
which a special class of sector nonlinearities is considered.\\

\emph{\textbf{Theorem 5.} Consider the Lipschitz nonlinear system
(\ref{sys3})-(\ref{sys4}) along with the observer (\ref{observer1}).
The observer error dynamics is (globally) asymptotically stable with
decay rate $\beta$ and simultaneously maximized admissible Lipschitz
constant $\gamma^{*}$ and minimized $\mathfrak{L}_{2}(w \rightarrow
e)$ gain, $\mu^{*}$, if there exist fixed scalers
$0\leq\lambda\leq1$ and $\beta>0$, scalers $\alpha>1$, $\epsilon>0$,
$\xi>0$ and $\zeta>0$ and matrices $P>0$ and $F$ such that the
following LMI optimization problem has a solution.}\\
\begin{equation}
\hspace{-5cm} \ min \ [\lambda\cdot\xi+(1-\lambda)\zeta] \notag
\end{equation}
\hspace{2cm} \emph{s.t.}
\begin{align}
&A^{T}P+PA+2\beta P-C^{T}F^{T}-FC < -\alpha I-\epsilon I\\
&\left[
 \begin{array}{cc}
    \frac{1-\bar{\sigma}^{2}(H)}{2}\cdot\xi I & P \\
    P &  \frac{1-\bar{\sigma}^{2}(H)}{2}\cdot\xi I
  \end{array}
 \right]>0\\
&\left[
  \begin{array}{ccc}
    H^{T}H+\frac{1}{2}(\xi-2\alpha)I & I & PB-FD \\
    I & -2\xi I & 0 \\
    B^{T}P-D^{T}F^{T} & 0 & -\zeta I \\
  \end{array}
\right]<0\label{LMI9}
\end{align}
\emph{Once the problem is solved,}
\begin{align}
L&=P^{-1}F\\
\gamma^{*}&\triangleq\max(\gamma)=\xi^{-1}\\
\mu^{*}&\triangleq\min(\mu)=\sqrt{\zeta}
\end{align}

\textbf{Proof:} The above is a scalarization of a multiobjective
optimization with two optimality criteria. Since each of these
optimization problems is convex, the scalarized problem is also
convex \cite{Boyd2}. The rest of the proof is the same as the proof
of Theorem 4 where the LMI \eqref{LMI9} is obtained from the LMI
\eqref{LMI3} using the Schur's complement lemma. \ $\triangle$


\section{Illustrative Examples}

In this section the high performance of the proposed observer is
shown via three design examples.\\

\hspace{0.5cm} \emph{\textbf{Example 1.}} Consider the following
observable (A,C) pair
\begin{eqnarray}
A=\left[%
\begin{array}{cc}
  0 & 1 \\
  1 & -1 \\
\end{array}%
\right], C=\left[
\begin{array}{cc}
  0 & 1
\end{array}
\right]\notag
\end{eqnarray}
The result of the iterative algorithm proposed in \cite{Rajamani} is
\begin{eqnarray}
\gamma^{*}&=&0.49\notag
\\L&=&\left[
\begin{array}{cc}
   69.5523 & 11.5679 \\
 \end{array}
 \right]^{T}\notag
\end{eqnarray}
while using our proposed method in Theorem 2,
\begin{eqnarray}
\gamma^{*}&=&1.1933\notag
\\L&=&\left[
\begin{array}{cc}
   56.8334 & 21.9074 \\
 \end{array}
 \right]^{T}\notag
\end{eqnarray}
which means that the admissible Lipschitz constant is improved by a
factor of $2.42$.\\

\begin{figure}[!h]
  \centering
  \includegraphics[width=6.5in]{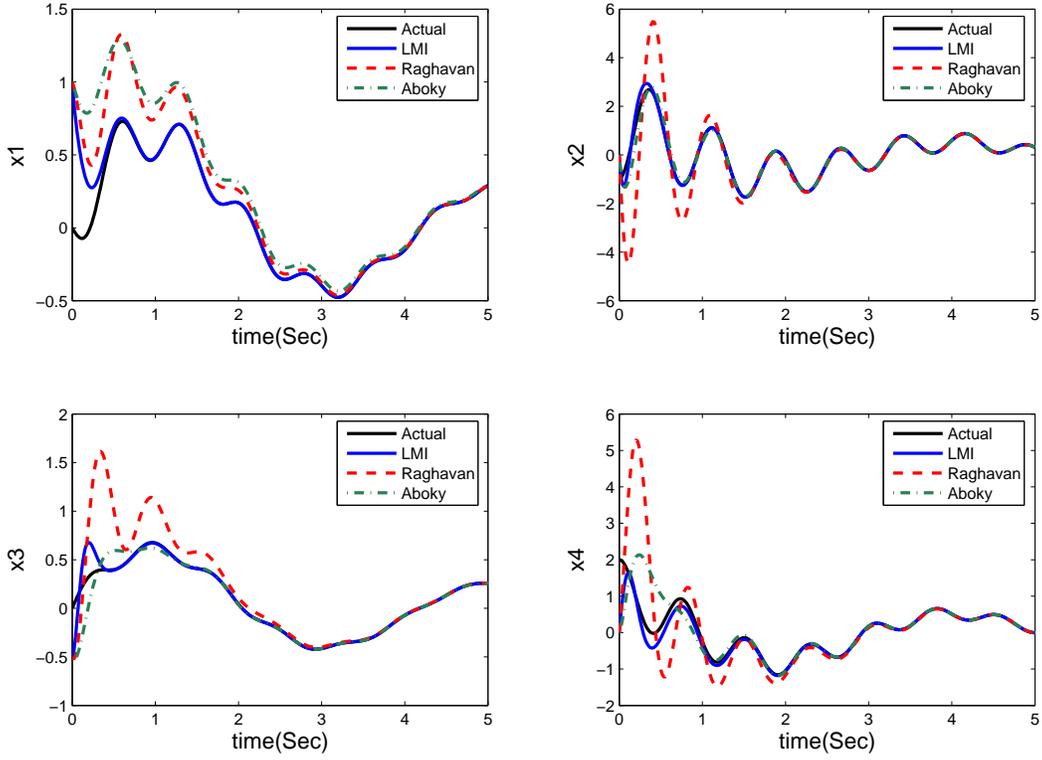}\\
  \caption{The true and estimated states of Example 2}\label{Fig1}
\end{figure}

\emph{\textbf{Example 2.}} The following system is the unforced
forth-order model of a flexible joint robotic arm as presented in
\cite{Raghavan}, \cite{Aboky}, \cite{Rajamani2}. The reason we have
chosen this example is that it is an important industrial
application and has been widely used as a benchmark system to
evaluate the performance of the observers designed for Lipschitz
nonlinear systems.
\begin{eqnarray}
\dot{x}&=&\left[
    \begin{array}{cccc}
      0 & 1 & 0 & 0 \\
      -48.6 & -1.25 & 48.6 & 0 \\
      0 & 0 & 0 & 1 \\
      19.5 & 0 & -19.5 & 0 \\
    \end{array}
  \right]x+\left[\begin{array}{c}
             0 \\
             0 \\
             0 \\
             -3.33\sin(x3)
           \end{array}\right]\notag\\
y&=&\left[
    \begin{array}{cccc}
      1 & 0 & 0 & 0 \\
      0 & 1 & 0 & 0 \\
    \end{array}
  \right]x.\notag
\end{eqnarray}
The system is globally Lipschitz with Lipschitz constant
$\gamma=3.33$. Noticing the structure of $\Phi$ that has a zero
entry in three of its channels, Raghavan \cite{Raghavan}, proposed
the coordinates transformation $\bar{x}=Tx$, where
\begin{equation}
T= diag \ [1,1,4,0.1]
\end{equation}
under which the transformed system has Lipschitz constant
$\bar{\gamma}=0.083$. Using Theorem 3, $\gamma^{*}=0.4472$ in the
original coordinates and $\bar{\gamma}^{*}=2.4177$ in the
transformed coordinates. The observer gain $\bar{L}$, is obtained in
the transformed coordinates and computed in the original coordinates
as $L=T^{-1}\bar{L}$. Assuming
\begin{eqnarray}
\beta&=&0.2\notag\\
B&=&\left[\begin{array}{cccc}
    1 & 1 & 1 & 1
  \end{array}\right]^{T}\notag\\
D&=&\left[
    \begin{array}{cc}
      0.1 & 0.25 \\
    \end{array}
  \right]^{T}\notag\\
H&=&0.5 I_{4\times4}\notag,
\end{eqnarray}\notag
using Theorem 4 we get, $\mu^{*}=0.5753$, $\alpha=2.0517$,
$\epsilon=0.0609$, and finally the observer gain will be
\begin{align}
L=\left[
\begin{array}{cccc}
  33.4865 & 129.9249 & 59.89713 & 108.2134 \\
  38.5694 & 282.8603 & 102.1561 & 171.0910
\end{array}\right]^{T}.\notag
\end{align}
Figure \ref{Fig1}, shows the true and estimated values of states.
The actual states are shown along with the estimates obtained using
Raghavan's and Aboky's methods and our proposed LMI optimization
method. The initial conditions for the system are $x(0)=\left[
                                              \begin{array}{cccc}
                                                0 & -1 & 0 & 2 \\
                                              \end{array}
                                            \right]^{T}$ and
those of the all observers are $\hat{x}(0)=\left[
                                              \begin{array}{cccc}
                                               1 & 0 & -0.5 & 0 \\
                                              \end{array}
                                            \right]^{T}$.
As seen in Figure \ref{Fig1}, the observer designed using the
proposed LMI optimization method has the best convergence of the
three. Note that in addition to the better convergence, the proposed
observer is an $H_{\infty}$ filter with maximized disturbance
attenuation level while the observers designed based on the methods
of \cite{Raghavan, Rajamani, Rajamani2, Aboky, Pertew} can only
guarantee stability of the observer error.
\\

\begin{figure}[!h]
  \centering
  \includegraphics[width=4.5in]{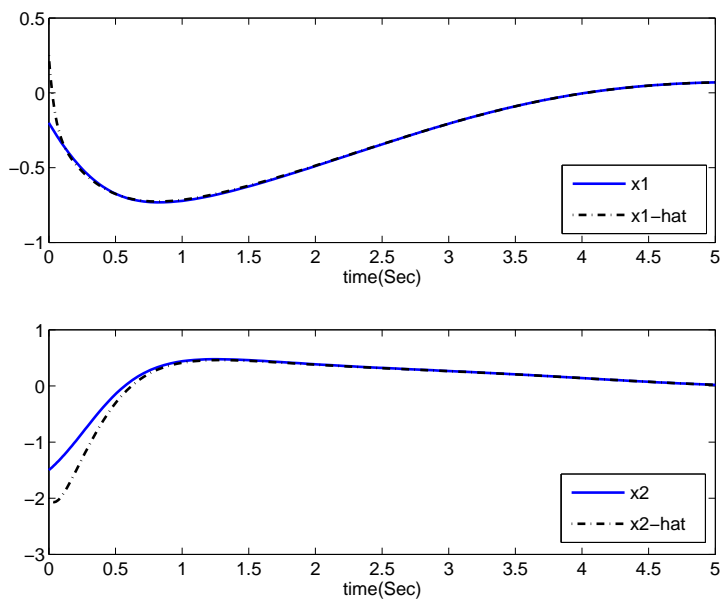}\\
  \caption{The true and estimated states of Example 3 in the presence of disturbance}\label{Fig2}
\end{figure}

\emph{\textbf{Example 3.}} In this example we show the usage of of
the multiobjective optimization of Theorem 5 in the design of
$H_{\infty}$ observers. Consider the following system
\begin{eqnarray}
x&=&\left[
    \begin{array}{cc}
    x_{1} & x_{2} \\
    \end{array}
\right]^{T}\notag \\
\dot{x}&=&\left[%
\begin{array}{cc}
  0 & 1 \\
  -1 & -1 \\
\end{array}%
\right]x
 + \left[
    \begin{array}{c}
        x_{1}^{3} \\
        -6x_{1}^{5}-6x_{1}^{2}x_{2}-2x_{1}^{4}-2x_{1}^{2} \\
    \end{array}\notag
\right] \\
y &=& \left[
\begin{array}{cc}
            1 & 0
          \end{array}
          \right]x.
\end{eqnarray}
The systems is locally Lipschitz. Its Lipschitz constant is
region-based. Suppose we consider the region $\mathcal{D}$ as
follows
\begin{eqnarray}
\mathcal{D}&=&\biggl\lbrace (x_{1},x_{2})\in \mathbb{R}^{2} \ | \
x_{1}\leq 0.25 \biggr\rbrace\notag
\end{eqnarray}
in which the Lipschitz constant is $\gamma=0.4167$. We choose
\begin{align}
H&=0.5I\notag \\
B&=\left[
       \begin{array}{cc}
       1 & 1 \\
       \end{array}
\right]^{T}\notag
\\D&=0.2\notag
\\\beta&=0.05\notag
\end{align}
and solve the multiobjective optimization problem of Theorem 5 with
$\lambda=0.9$. We get
\begin{align}
\gamma^{*}&=0.5525\notag\\
\mu^{*}&=1.1705\notag\\
\alpha&=1.6260\notag\\
\epsilon&=2.2435 \times 10^{-4}\notag\\
L&=\left[
\begin{array}{cc}
   23.7025 & 13.7272 \\
\end{array}
\right]^{T}.\notag
\end{align}
The true and estimated values of states are shown in Figure
\ref{Fig2}. We have assumed that
\begin{eqnarray}
x(0)&=&\left[\begin{array}{cc}
       -0.2 & -1.45
     \end{array}\right]^{T} \notag \\
\hat{x}(0)&=&\left[\begin{array}{cc}
       0.25 & -2
     \end{array}\right]^{T}\notag
\\w(t)&=&0.15\exp(-t)\sin(t).\notag
\end{eqnarray}
%

\begin{figure}[!h]
  \centering
  \includegraphics[width=5.5in]{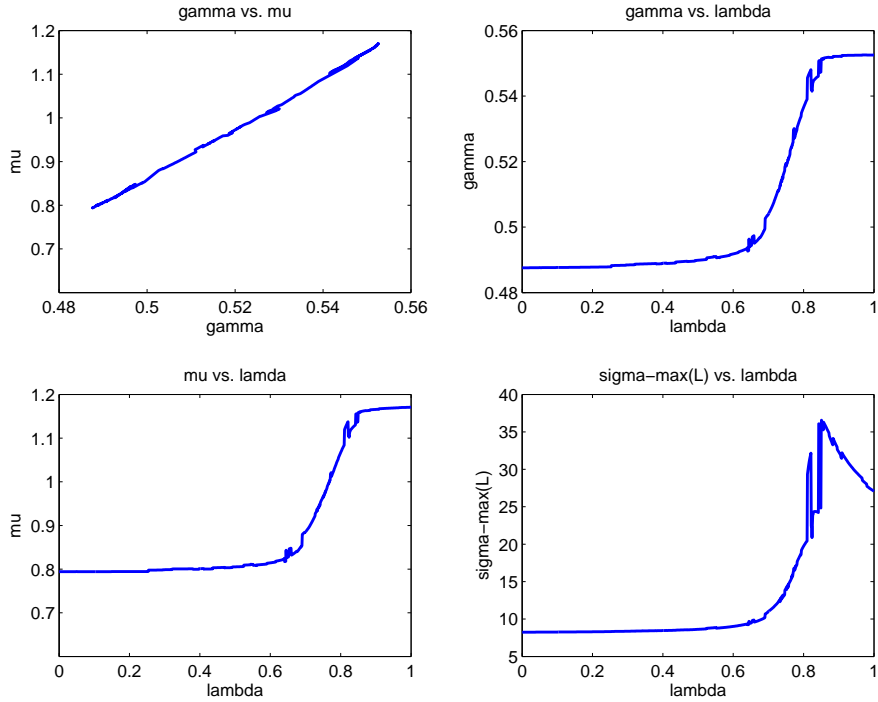}\\
  \caption{$\gamma^{*}$, $\mu^{*}$ and
$\bar{\sigma}(L)$, and the optimal trade-off curve with
$\beta=0.05$}\label{Fig3}
\end{figure}

\begin{figure}[!h]
  \centering
  \includegraphics[width=5.5in]{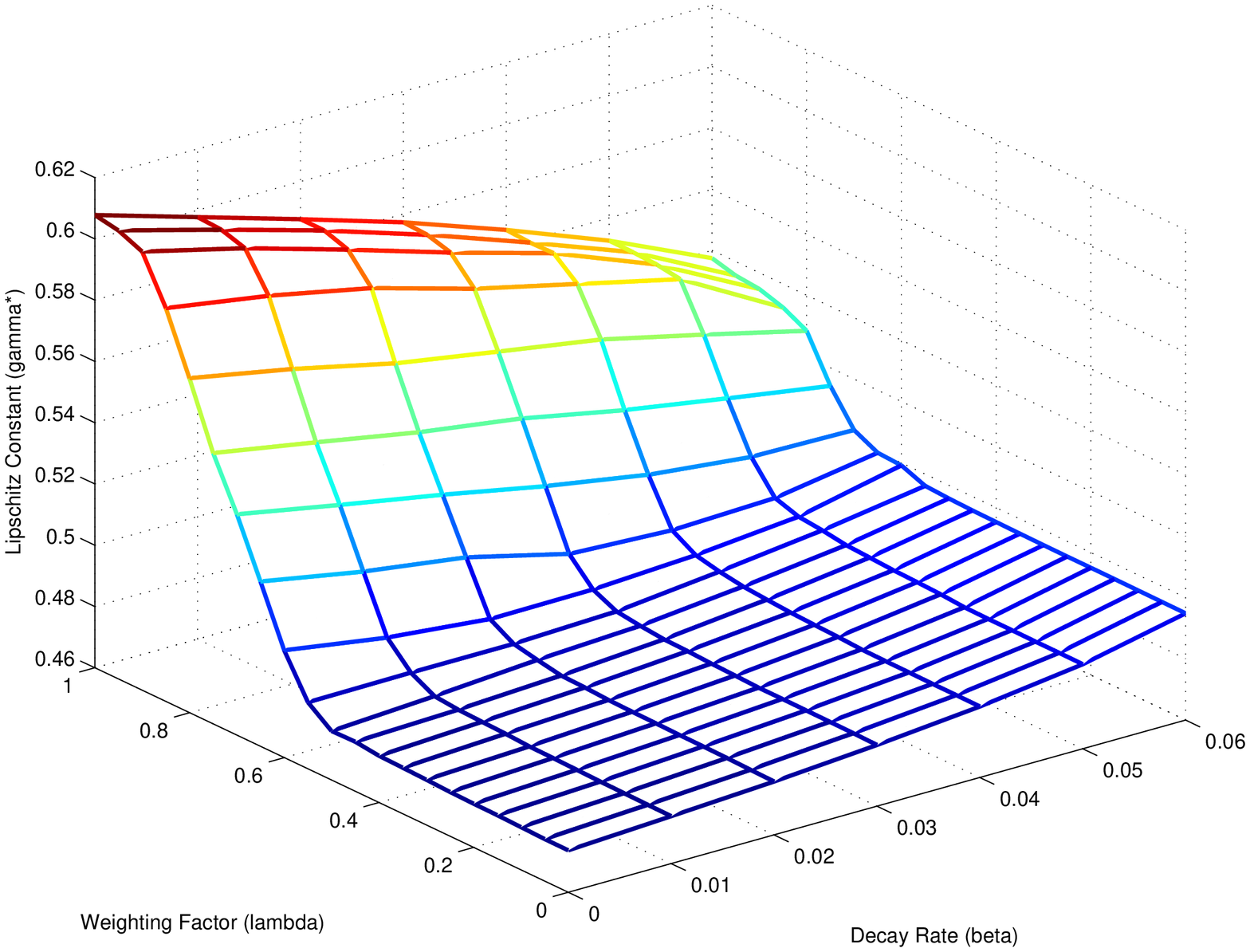}\\
  \caption{The optimal surface of $\gamma^{*}$}\label{Fig4}
\end{figure}

\begin{figure}[!h]
  \centering
  \includegraphics[width=5.5in]{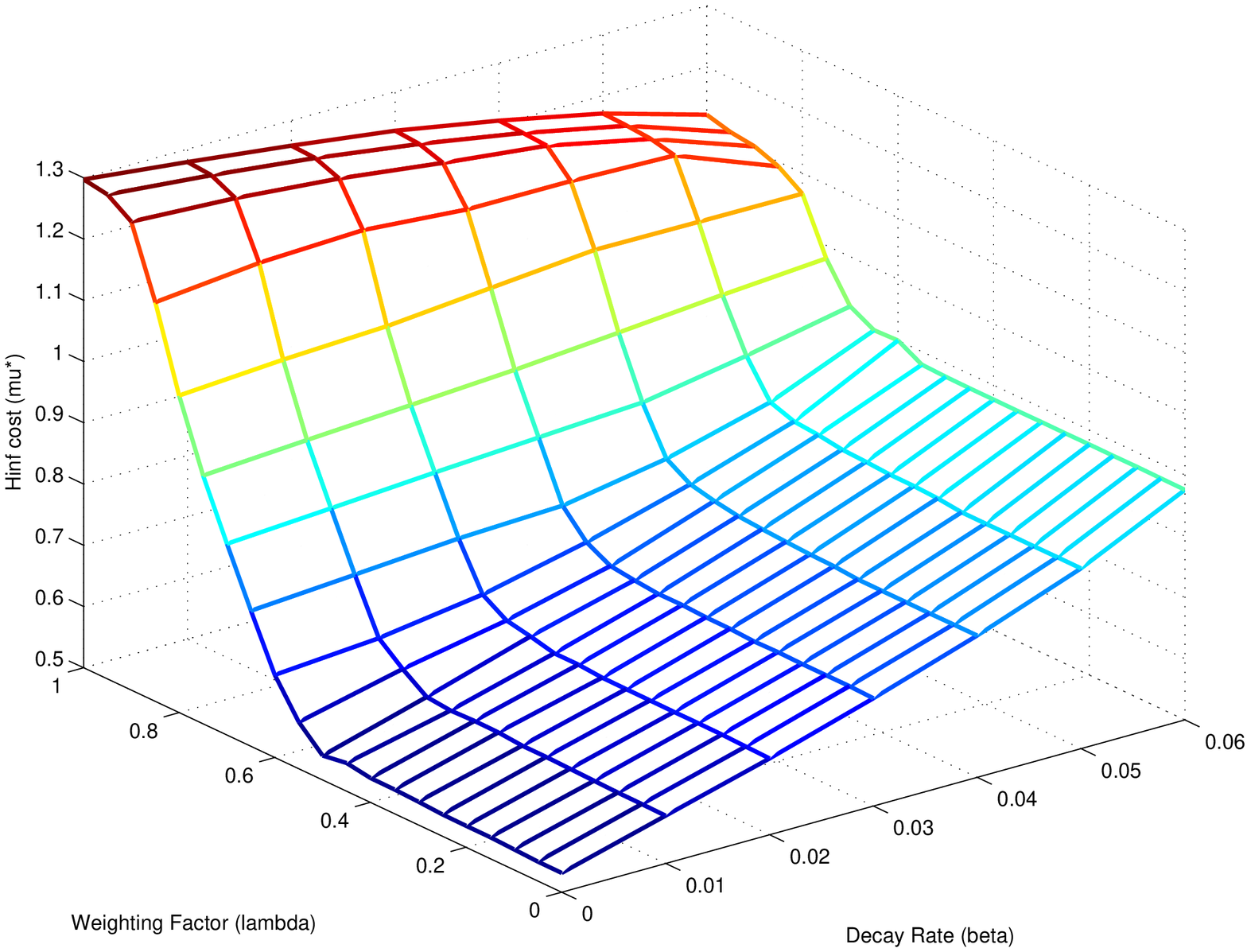}\\
  \caption{The optimal surface of $\mu^{*}$}\label{Fig5}
\end{figure}

The values of $\gamma^{*}$, $\mu^{*}$, norm of the observer gain
matrix, $\bar{\sigma}(L)$, and the optimal trade-off curve between
$\gamma^{*}$ and $\mu^{*}$ over the range of $\lambda$ when the
decay rate is fixed ($\beta=0.05$) are shown in Figure \ref{Fig3}.
The optimal surfaces of $\gamma^{*}$ and $\mu^{*}$ over the range of
$\lambda$ when the decay rate is variable are shown in Figures
\ref{Fig4} and \ref{Fig5}, respectively.

\section{Conclusions}

A new method of robust observer design for Lipschitz nonlinear
systems proposed based on LMI optimization. The Lipschitz constant
of the nonlinear system can be maximized so that the observer error
dynamics not only be asymptotically stable but also the observer can
tolerate some additive nonlinear uncertainty. In addition, the
result extended to a robust $H_{\infty}$ nonlinear observer. The
obtained observer has three features, simultaneously. Asymptotic
stability, robustness against nonlinear uncertainty and minimized
guaranteed $H_{\infty}$ cost. Thanks to the linearity of the
proposed LMIs in both admissible Lipschitz constant and the
disturbance attenuation level, they can be simultaneously optimized
through convex multiobjective optimization. The observer high
performance showed through design examples.



\bibliographystyle{IEEEtran}
\bibliography{Thesis_References}






\end{document}